\documentclass[sigconf]{acmart}
\usepackage{subfig}

\usepackage{pgf, tikz}

\usepackage{algorithm} 
\usepackage{algpseudocode} 
\usepackage{listings}
\usepackage{hyperref}
\usepackage{enumitem}
\usepackage{tabularx}
\usepackage{array}
\makeatletter
\g@addto@macro{\endtabular}{\rowfont{}}
\makeatother
\newcommand{\rowfonttype}{}
\newcommand{\rowfont}[1]{
   \gdef\rowfonttype{#1}#1%
}
\newcolumntype{Y}{>{\centering\arraybackslash\rowfonttype}X}

\AtBeginDocument{%
  \providecommand\BibTeX{{%
    \normalfont B\kern-0.5em{\scshape i\kern-0.25em b}\kern-0.8em\TeX}}}

\setcopyright{acmcopyright}
\copyrightyear{2020}
\acmYear{2020}

\acmConference[CIDR '21]{CIDR '21: Conference on Innovative Data Systems Research}{January 12--15, 2021}{CIDR, Amsterdam}
\acmBooktitle{CIDR '21: Conference on Innovative Data Systems Research, January 12--15, 2021, CIDR, Amsterdam}
\acmISBN{978-1-4503-XXXX-X/18/06}

\begin{document}

\title{Railgun: streaming windows for mission critical systems}


\author{Jo\~ao Oliveirinha}
\email{joao.oliveirinha@feedzai.com}
\affiliation{%
  \institution{Feedzai}
}

\author{Ana Sofia Gomes}
\email{sofia.gomes@feedzai.com}
\affiliation{%
  \institution{Feedzai}
}

\author{Pedro Cardoso}
\email{pedro.cardoso@feedzai.com}
\affiliation{%
  \institution{Feedzai}
}

\author{Pedro Bizarro}
\email{pedro.bizarro@feedzai.com}
\affiliation{%
  \institution{Feedzai}
}

\renewcommand{\shortauthors}{Oliveirinha, et al.}

\begin{abstract}
Some mission critical systems, such as fraud detection, require accurate, real-time metrics over long time windows on applications that demand high throughputs and low latencies. 
As these applications need to run ``forever'', cope with large and spiky data loads, they further require to be run in a distributed setting.
Unsurprisingly, we are unaware of any distributed streaming system that provides all those properties. Instead, existing systems take large simplifications, such as implementing sliding windows as a fixed set of partially overlapping windows, jeopardizing metric accuracy (violating financial regulator rules) or latency (breaching service agreements).
In this paper, we propose Railgun, a fault-tolerant, elastic, and distributed streaming system supporting real-time sliding windows for scenarios requiring high loads and millisecond-level latencies. We benchmarked an initial prototype of Railgun using real data, showing significant lower latency than Flink, and low memory usage, independent of window size.
\end{abstract}

\ccsdesc[500]{Computer systems organization~Client-server architectures}
\ccsdesc[500]{Computing methodologies~Distributed algorithms}

\keywords{distributed systems, streaming engines, fraud detection systems}

\maketitle

\section{Introduction}
\label{sec:introduction}
In some low-latency mission critical systems, such as in financial fraud
detection, where transactions require a decision in a few milliseconds, it is desirable that the underlying streaming engines fulfill all the following L-A-D requirements:
\begin{itemize}
  \item[] \textbf{L}ow latencies even at high percentiles (<250ms @ 99.9\%);
  \item[] \textbf{A}ccurate metrics event-by-event;
  \item[] \textbf{D}istributed, scalable and fault-tolerant.
\end{itemize}

However, to the best of our knowledge, no streaming engine delivers all three LAD requirements. Type 1 streaming engines such as  Aurora~\cite{DBLP:conf/sigmod/AbadiCCCCEGHMRSSTXYZ03}, STREAM~\cite{DBLP:conf/sigmod/ArasuBBDIRW03}, TelegraphCQ~\cite{DBLP:conf/cidr/ChandrasekaranDFHHKMRRS03} and others \cite{suhothayan2011siddhi,DBLP:conf/sigmod/ChenJDTW00,trill} provide low latency and accurate results but do not scale beyond one node (LA). Type 2 engines such as Flink~\cite{DBLP:journals/debu/CarboneKEMHT15}, Kafka,  Streams~\cite{kreps2016introducing}, and others \cite{10.1145/3183713.3190664, Akka,noghabi2017samza,ramasamy2019unifying} provide scalability and fault-tolerance (LD) at the expense of inaccurate results due to window choices (see below) or load shedding~\cite{DBLP:conf/cidr/AbadiABCCHLMRRTXZ05}.
\vspace{-0.25cm}
\begin{table}[h]
    \centering
    \begin{tabularx}{0.4\textwidth}{lYYY}
        \cline{2-4}
        \rowfont{\large}&\textbf{L} &\textbf{A} &\textbf{D} \\
        \rowfont{\scriptsize}&Low latency at high percentiles &Accurate metrics event-by-event &Distributed, scalable and fault-tolerant \\
        \hline
        \rowfont{\normalsize}Type 1 engines &Yes &Yes &No \\
        Type 2 engines &Yes &No &Yes \\
        \textbf{Railgun} &Yes &Yes &Yes\\
        \hline
    \end{tabularx}
    \caption{Options taken by different streaming engines}
    \label{tab:lad}
\end{table}
\vspace{-0.7cm}

A critical decision is how to handle a large streaming state while delivering low latency. In low throughputs and small windows, events can fit in-memory of a single node, and accurate metrics can be computed for every new event. However, for large windows or high throughputs (i.e., where \textbf{D} is required) handling the incoming \emph{and} expiring events becomes such a problem that Type 2 systems either shed load, or use hopping windows as an approximation of real sliding windows, expiring events only so often. For instance, a 5-min sliding window can be approximated, e.g., using five fixed 5-min windows, offset-ed by 1 minute (the \emph{hop}) and where, as time passes, new windows (and their aggregations) are created and expired.
\begin{figure}[b]
    \centering
    \includegraphics[scale=0.7]{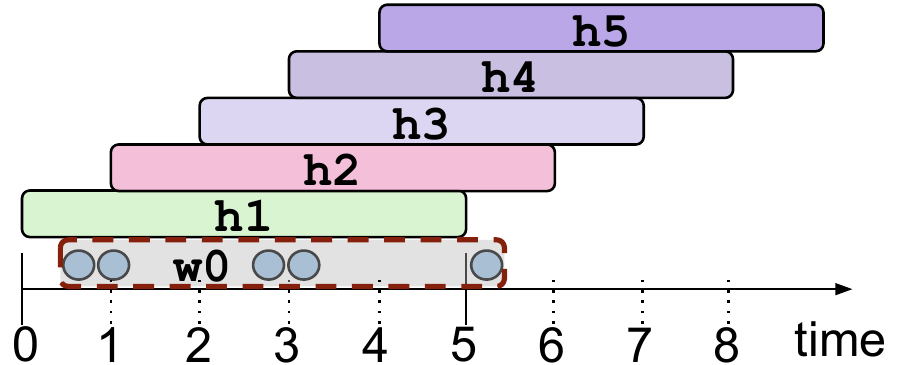}
    \caption{A 5-min hopping window with 1-min hop uses physical windows h1-h5 but none capture the 5 events (circles), unlike a real-time sliding window (w0).}
    \label{fig:hop-example}
\end{figure}
As seen in Figure~\ref{fig:hop-example}, approximating sliding windows with hoping windows can lead to inaccurate aggregations: a true sliding window can count 5 events, but a hopping window with 1-min hop might not. The hop could be smaller, e.g., 1 second, but that would imply concurrently managing 300 5-min windows, instead of 5. 

This state of affairs presents a challenge for modern fraud detection systems. Responsible for processing trillions of dollars worldwide every year, these are mission-critical systems with demanding latency requirements (e.g., <250 ms for  99.9\% percentile), while still requiring accurate metrics event-by-event (for regulatory and adversarial reasons, see Section~\ref{sec:realtimewindows}).
To address this need, we propose Railgun, a novel distributed streaming engine based on low-memory-footprint, disk-backed sliding windows (an improvement on the sliding windows and SlideM algorithm~\cite{slidem}) on top of which, we built state-management and distributed communication layers to fulfill the LAD requirements. Our contributions are as follows:
\begin{enumerate}
    \item We formulate the problem of using hopping windows in use-cases such as fraud-detection (Section~\ref{sec:realtimewindows});
    \item We present our proposal, Railgun, with an overview of the architectural design, components and decisions (Section~\ref{sec:railgun});
    \item We show that Flink performance degrades when small hops are used to approximate real sliding windows (Section~\ref{subsection:flinkVsRailgun});
    \item We show that Railgun computes real-time metrics over large windows in an efficient and scalable way (Section~\ref{subsec:scalingRailgun}).
\end{enumerate}

\section{Background}
\label{sec:background}
A data stream $S$ is an unbounded sequence of events $e_1, e_2, ...$, each a data point with a timestamp. Aggregations over streams are computed using windows. A \emph{window} $w$ is a sequence of contiguous data over $S$ with a certain size $w_s$ (defined by a number of events, time interval, or start-stop conditions as in a user session). In this paper, we focus on time-based windows, henceforth referred simply as windows.
As time passes, a window over a stream is evaluated often, at a specific, and changing, timepoint $T_{eval}$. $T_{eval}$ determines the events to include for the aggregations, where an event with timestamp $t_i$ belongs to a window evaluation iff $T_{eval} - w_s \leq t_i < T_{eval}$.

\emph{Hopping windows} are windows where $T_{eval}$ changes according to a step of length $s$.
This step $s$, or \emph{hop}, marks \emph{when} new windows are created. If $s$ is smaller than $w_s$, then the windows overlap, i.e., an event may belong to more than one hopping window\footnote{Hopping windows are often called sliding windows by systems such as Flink because they \emph{approximate} the behavior of real sliding windows.}. When $s$ is equal to $w_s$, hopping windows do not overlap, and events belong to exactly one window. This case is frequently given the name of \emph{tumbling windows}. Step $s$ is generally not bigger than $w_s$.

\emph{Real sliding windows}, or just \emph{sliding windows}, are windows where $T_{eval}$ is the moment right after a new event has arrived. This frequent evaluation is computationally expensive as, for each new event $e_i$, the system has to expire events and (re-)compute aggregations, but on the other hand, aggregations are always accurate.

\subsection{Fraud-detection requires LAD}
\label{sec:realtimewindows}
%
%

While for most streaming use-cases hopping windows are enough, fraud-detection requires all LAD properties, as shown next.

Fraud-detection systems make decisions over financial transactions, e.g., by blocking a transaction, or raising an alarm when money laundering is suspected. Such systems use streaming aggregations as inputs for models and rules to make decisions~\cite{kdd-Branco-deeplearning}. For instance, queries such as \verb|Q1| and \verb|Q2| below can be used to profile cards or merchants, and detect suspicious behavior:
\begin{lstlisting}[
           language=SQL,
           showspaces=false,
           basicstyle=\ttfamily\small,
           commentstyle=\color{gray},
           label={lst:query},
           caption={Streaming 5-min metrics per card and merchant.},
           captionpos=b
        ]
Q1: SELECT SUM(amount), COUNT(*) FROM payments 
    GROUP BY card [RANGE 5 MINUTES]
Q2: SELECT AVG(amount) FROM payments
    GROUP BY merchant [RANGE 5 MINUTES]
\end{lstlisting}

%
Profiles computed over hopping windows are weaker as they are vulnerable to adversary attacks. E.g., fraudsters can schedule attacks to occur on specific times, or follow a specific cadence, taking advantage of the predictable hop size. In addition, since those profiles can be inaccurate, they compromise rule compliance -- either from internal by-laws, or from external regulators. 
As illustration, consider the following business rule: “\emph{if} the number of transactions of a card in 5 minutes is higher than 4, \emph{then} block the transaction”. If the window is implemented using 1-min hops, then the situation in Figure~\ref{fig:hop-example} can happen: the rule should trigger on the fifth event as it arrives within 5 minutes of the first event, but there is no hopping window including all 5 events in its boundaries using a 1-min hop.

Failing to ensure precision over these rules, may cause penalties, heavy sanctions and reputational damage. 
To avoid this, one could argue that the hop could be adjusted to catch the intended behavior. However, the solution is not a panacea. First, the problem in Figure~\ref{fig:hop-example} can happen regardless of size. Second, if the hop is reduced to a size where hopping windows behave almost like real sliding windows (e.g., 1-ms step) then most stream processing engines systems crash or significantly degrade performance (cf. Section~\ref{subsection:flinkVsRailgun}).
This problem worsens with long windows. Fraud profiles use windows spanning over days, weeks, months and sometimes years. These include, e.g., the number of distinct addresses used in the last 6 months, or
the average user's expenditure of the past year.

Fraud-detection systems have very strict and demanding requirements. Namely they must: 1) provide sub 250ms latency over 99.9\% percentile (\textbf{L}); 2) provide accurate metrics per-event (\textbf{A}); 3) be distributed, horizontally scalable and fault-tolerant (\textbf{D}).

\begin{figure*}[ht!]
    \centering
      \includegraphics[width=0.90\textwidth]{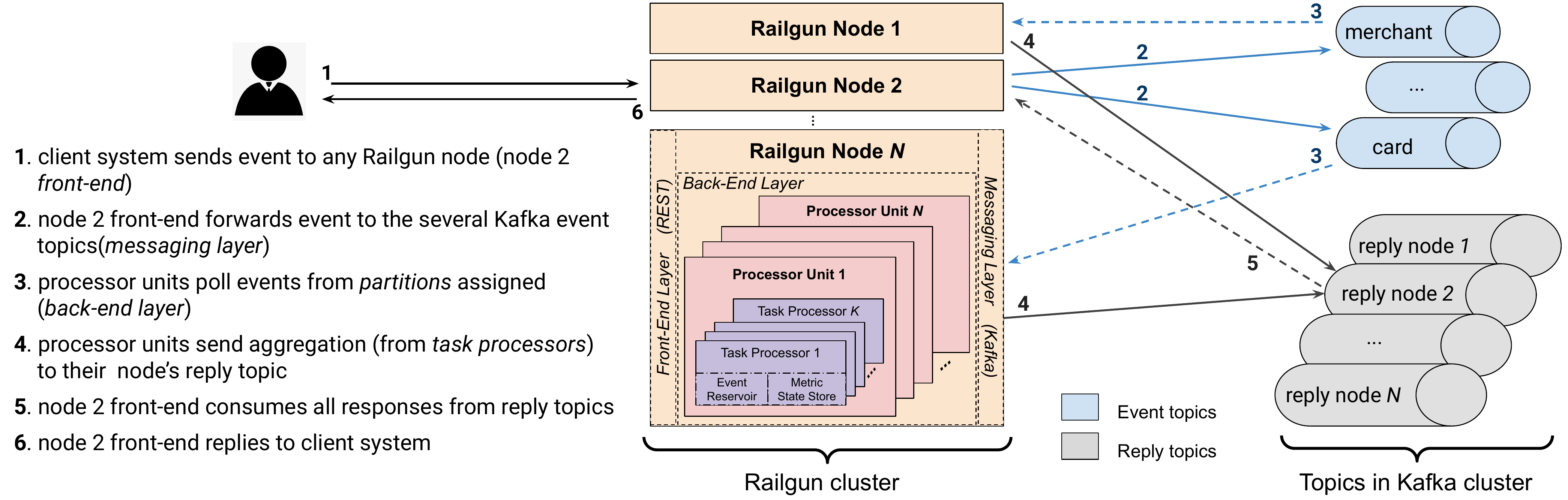}
    \caption{Tale of an event traversing Railgun.}
    \label{fig:tale-event}
\end{figure*}

\subsection{Distributed Streaming Engines}
\label{subsec:stateofart}
To the best of our knowledge, no known, distributed streaming engine uses real sliding windows but instead implements hopping windows, thereby failing to meet requirement \textbf{A}, i.e., accurate metrics per event. This happens in Flink, Kafka Streams, Spark Streaming and others, which we typify as Type 2 systems (Table~\ref{tab:lad}).


Besides the functional limitations shown in Section~\ref{sec:realtimewindows}, there are also non-functional drawbacks of hopping windows with respect to latency, CPU usage and state scalability. So why are they used?
%
Hopping windows can avoid storing events. 
Since $s$ and $w_s$ do not change during run-time, the number of physical window states is fixed and exactly $\frac{\texttt{windowSize}}{\texttt{hopSize}}$. Arriving events update those window states, but can be discarded once its contribution has been applied. Hence, besides saving storage, these solutions also avoid processing event expiration.
Recall \verb|Q1| and Figure~\ref{fig:hop-example}: \verb|sum| and \verb|count| payments made by a card in the last 5 min with 1-min hop. Any event for this window affects 5 window states, and in this case, two variables per window state. Every minute and, for every card active in the last 5 min, new variables are created and the oldest, expired.
%
%
This property makes hopping windows interesting when the ratio $\frac{\texttt{windowSize}}{\texttt{hopSize}}$ is low, as it is independent of throughput. When the ratio is higher, hopping windows bring extra problems (cf. Section~ \ref{subsection:flinkVsRailgun}).

%

Despite these drawbacks, hopping windows' wide usage and characteristics have driven substantial research and optimizations such as Cutty~\cite{DBLP:conf/cikm/CarboneTKHM16}, Scotty~\cite{DBLP:conf/icde/TraubGCBKRM18} and others~\cite{DBLP:conf/sigmod/TheodorakisKPP20, DBLP:conf/edbt/BensonGZMR20} 
that contribute to its popularity by delivering reduced hardware costs, support for out-of-order events, and distributed computations of a single query.

Flink acknowledges the issue of high-precision metrics over time windows for low latency fraud-detection, with a customized window solution on-top of the code-base to address it~\cite{flink-fraud}. However, this approach  has quadratic performance. For each event, it  computes the aggregation from scratch by iterating over all stored events (persisted in RocksDB) for those matching the window interval. Since events are not first-class citizens, few optimizations are possible and performance worsens with long windows -- RocksDB is not designed for the FIFO streaming event access pattern, and the many repeated calculations make the solution fail requirement \textbf{L}.


\section{Railgun}
\label{sec:railgun}

Railgun is the paper's main contribution, and takes different design decisions when compared to the alternatives above: 1) it works with real sliding windows to achieve aggregation correctness at all times (and not just at every hop); 2) it uses an \emph{event reservoir} to efficiently store and access events; 3) it manages an external aggregation store (persisted in RocksDB) and a messaging layer (using Kafka) for distributed processing, fault-tolerance and checkpointing. 
%
The reservoir exploits the predictable, time access of events to optimize transfers between memory and stable storage, and access nearly all events from memory using an eager caching. 
Plus, by optimizing the computation and storage of aggregation states, Railgun can deliver accurate results, per-event, with low latency.


Railgun's high-level architecture is presented in Figure~\ref{fig:tale-event}, where we show a new event traversing the system. 
At this stage, and to simplify development, all Railgun nodes are equal and composed by \emph{messaging}, \emph{front-end}, and \emph{back-end} layers; described next. This could be revised in the future, with different nodes split by function.


\subsection{Messaging Layer}
The goal of the messaging layer is twofold: 1) serve as the communication layer between different Railgun nodes, including between the front-end and back-end layer of the same physical node; 2) support the recovery of Railgun node failures, by reliably storing events and aggregation replies which can be rewinded upon request. 

Concretely, Railgun uses Kafka~\cite{kreps2011kafka}. Kafka is a distributed, highly scalable and fault tolerant messaging system, with high throughput and low latency guarantees. In opposition to push-based systems such as RabbitMQ~\cite{rabbitmq}, Kafka follows a pull-based approach where consumers continuously poll for new messages by providing their individual offset since the last poll. This is important since it allows a Railgun node to recover by rewinding the stream and replaying unprocessed messages without degrading the end-to-end latency of the overall system.
Kafka stores messages in \emph{topics} and provides built-in capabilities to split topics into several \emph{partitions}. As we shall see in Section~\ref{subsec:backend}, Railgun exploits Kafka's partitions to manage work distribution between several Railgun processor units. 
However, and although we use Kafka as the messaging layer implementation, we could have chosen other systems such as Pulsar~\cite{pulsar}, as long as they implement the same pull-based and partition concepts.

\subsection{Front-End Layer}
The front-end layer is the entry point for client requests, including events, requests for new metrics/streams, or deletions. 
Hence, besides communicating with the client, the front-end layer distributes events and manages the overall cluster state (both using Kafka).

When a new stream is registered by the client, the front-end creates a set of partitioned topics to support it. The number of topics needed per stream depends on the set of metrics configured, and more precisely, on the number of distinct \emph{group by} fields of the stream. 
Recall Example~\ref{lst:query}. Since Railgun is distributed across several processing units, to have accurate metrics, we need to ensure that the processing unit computing metrics for a specific entity (e.g., card or merchant), receives all events from that entity. 
This is done by a routing task in the front-end, where events are hashed based on their known \emph{group by} keys. Multiple metrics over a stream, as in \verb|Q1| and \verb|Q2| of Example~\ref{lst:query}, cause the event to be forwarded to more than one topic (step 2 of Figure~\ref{fig:tale-event}).
Events are, in fact, replicated as many times as the number of top-level entities needed for a stream, resulting in different topics per stream.
Still, the number of topics needed is usually a small number, and it is not necessarily equal to the number of distinct \emph{group by} keys of all stream metrics (which could lead to dozens of topics).
Accurate metrics, only need events to be hashed by a \emph{subset} of their group by keys.
E.g., two metrics, one grouping by \texttt{card} and \texttt{merchant}, and the other by \texttt{card}, could both use topic \texttt{card}. This reduces Kafka's storage, and thus requires the front-end to receive, from configuration, the topics per stream. 
%

Metrics of a stream might be computed by multiple back-end instances residing in different Railgun nodes. Hence, the front-end is also responsible for collecting the several computations (step 5) from its reply topic, and responding to the client (step 6). 

\subsection{Back-End Layer}
\label{subsec:backend}
This is where the actual computation of metrics takes place. 
Within a back-end instance, we can have a number of configured \emph{processor units}. Each processor unit has its own dedicated thread, and is responsible for a subset of the topic partitions of multiple streams. By using a single thread, we reduce context switching and synchronization, thereby optimizing for event latency.
The number of processor units, among the several Railgun nodes, establishes the level of parallelism within the cluster.
Importantly, two processor units on the same Railgun node are equivalent to two Railgun nodes with one processor unit each. By having many processor units inside a single node, we can exploit multi-core machines efficiently.
%
\vspace{-0.4cm}
\begin{algorithm}
	\caption{Processor Unit Logical Loop} 
	\label{algo:processor}
	\begin{algorithmic}[1]
		\While {$running$}
		    \State Check for operational $tasks$ and process them
		    
			\State $messages \leftarrow consumer.poll(timeout)$
			
			\For {$message \leftarrow messages$}
			    \State $t \leftarrow message.topic$
			    \State $p \leftarrow message.partition$
			    \State $taskProcessor \leftarrow taskProcessors.get(t, p)$
                \State $taskProcessor.processMessage(message)$
			\EndFor
		\EndWhile
	\end{algorithmic} 
\end{algorithm}
\vspace{-0.4cm}

Algorithm~\ref{algo:processor} summarizes the processor unit's duties. While running, the processor checks for operational tasks such as adding/removing new streams or metrics; and forwards events for the actual computation of metrics to the appropriate task processors.

Each \emph{task processor} handles a single pair \emph{(topic, partition)}. I.e., a processor will have as many task processors as \emph{(topic, partition)} pairs assigned. There is only one active task processor for each \emph{(topic, partition)} pair in the whole cluster, and thus the combination of unique topic-partition pairs sets the cluster's level of concurrency.
Most of the back-end's complexity, such as topic-partition assignment, is delegated to the messaging layer. When a node fails, the messaging layer detects the failure and, at step 3 of Algorithm~\ref{algo:processor}, triggers a callback to assign these partitions to another processor. 

Each task processor is composed of: an \textit{event reservoir} that stores the events; a \textit{state store} with aggregation states of each configured metric; and a \textit{plan} -- directed acyclic graph (DAG) of operators. 

\subsubsection{Event Reservoir}
\label{subsec:reservoir}
Processing an event starts with the event reservoir, where events are persisted to disk.
In it, events are serialized and compressed into groups of contiguous \emph{chunks}. We do this to reduce the number of I/O operations, all of which are asynchronous, so as to not affect event processing latency. Chunks hold multiple events and are kept in-memory until they reach a fixed size, after which they are persisted to disk over \emph{immutable} and \emph{ordered} files, to support efficient random reads of events. 
By persisting chunks to disk often, we ease recovery, as only the most recent events can be lost, and recovered from the Messaging Layer.

Since events are consumed in order, by advancing windows, the reservoir is able to provide \emph{iterators} that transparently load chunks into memory. Furthermore, iterators eagerly load adjacent chunks into cache when a new chunk is loaded from disk, and starts to be iterated. As a result, when a window needs events from the next chunk, it is normally already available for iteration. This predictability helps us relax the hardware demands for the reservoir, as even for low latency scenarios, we can use a network-attached storage or HDDs, reducing the total cost of ownership.
This reservoir is an evolution of previous work~\cite{slidem}. We use a locally-attached storage in each Railgun node to avoid a single point of failure, and define a data format and compression for efficient storage, both in terms of deserialization time and size. The latter is important since events can be replicated across multiple task processors.

\begin{figure}[h]
    \centering
    \includegraphics[width=0.45\textwidth]{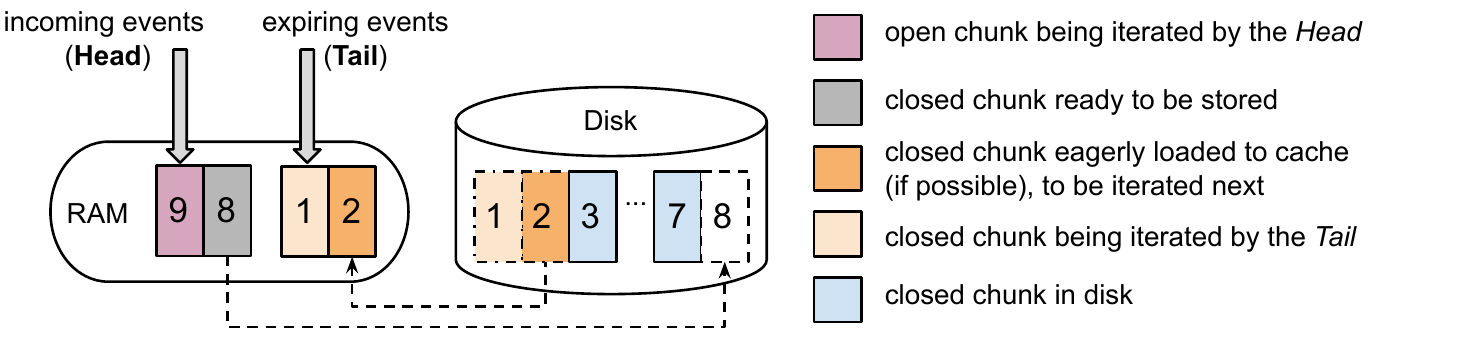}
    \caption{Iterators for a Window in Event Reservoir}
    \label{fig:reservoir}
\end{figure}

%

%
As illustrated in Figure~\ref{fig:reservoir}, by default, each window has two iterators - one for the head of the window (expiring events), and another for the tail (arriving events) - and each iterator only needs one chunk in-memory\footnote{However, due to eager caching, more chunks may be in-memory.}. Whenever possible, we reuse iterators among windows. Namely, over the same reservoir, two real-time sliding windows always share the same tail iterator (e.g. a 1-min and a 5-min window share the same tail iterator, which points to the most recently arrived event). 
%
This design makes the reservoir optimal for I/O~\cite{DBLP:journals/ipl/Roy07}, and extremely efficient for long windows. Namely, and except for the extra storage needed (minimized by compression and serialization), windows of years are equivalent to windows of seconds -- in performance, accuracy, and memory consumption.

\subsubsection{Plan and State Store}

The plan is a DAG of operations that compute all the metrics within a topic-partition, following the order: \texttt{Window -> Filter -> Group By -> Aggregator}. Since
we often have metrics sharing the \texttt{Window}, \texttt{Filter}, and \texttt{Group By} operators, the plan optimizes these by reusing the DAG's prefix path.

\begin{figure}[hb]
    \centering
    \includegraphics[width=0.35\textwidth]{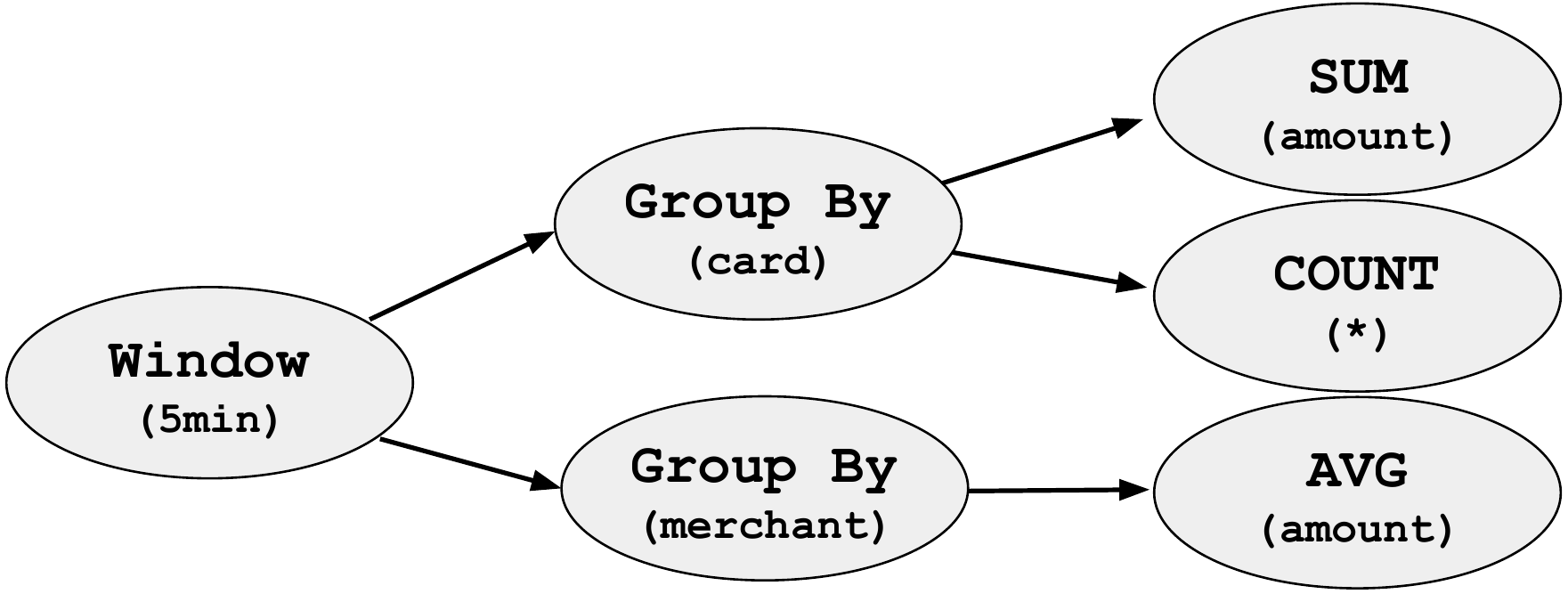}
    \caption{Plan DAG of Example~\ref{lst:query}.}
    \label{fig:dag}
\end{figure}

Figure~\ref{fig:dag} shows the DAG of Example~\ref{lst:query}. In it, all metrics share the same window, but \verb|Q1| groups by field \texttt{card} while \verb|Q2| by field \texttt{merchant}.
Optimizing the DAG to reuse operators prevents us from repeating unnecessary computations, especially ones related with windows. Every time a plan advances time, the \texttt{Window} operator produces the events that \emph{arrive} and \emph{expire}, to the downstream operators of the DAG. 
However, to make these optimizations, we restrict Railgun's query expressibility to follow a strict order of operations. In contrast, general solutions like Flink or Spark Streaming provide a more flexible API, which make it harder to optimize.

While the roots of the DAG iterate over the reservoir and push events downstream, the leafs (i.e., \texttt{Aggregator} operators) use the \textit{state store} to keep and access the results of the aggregations. Similarly to Flink, Railgun uses RocksDB for this purpose which has proven to be a reliable and low latency embedded key-value store.

\section{Experiments}
To validate our approach we present two experiments: 1) measure how Railgun's real-time sliding windows compare with Flink's hopping windows; 2) assess how Railgun scales with window size or number of windows. We chose Flink as it is one of the most used stream processing system, and the closest to our functional needs. Space constraints prevent us from showing how Railgun scales to address higher throughputs in multi-node setup. However, note that in Railgun's architecture (Section~\ref{subsec:backend}) having several task processors (due to topic partition) within a single node is equivalent to multiple nodes with a single task processor; and since Railgun is under development, assessing performance and scalability within a single node already allow us to validate several design decisions.

\subsection{Setup and Methodology}
In all experiments we use 3 \textit{m5.2xlarge} AWS instances\footnote{https://aws.amazon.com/ec2/instance-types/}, using only EBS storage. We use Kubernetes to deploy 1 Kafka \emph{pod} (with Zookeeper), 1 injector, and 1 computing engine -- either Railgun or Flink (v1.11.0) -- with a JVM heap of 10GB all in separate VMs.
%
The injector sends events to a Kafka event topic, at a sustained throughput of 500 ev/s. 
The goal is to validate how latencies vary between engines with the same throughput and cluster setup.

Each computing engine consumes events from the input topic, computes the aggregation, and sends the result to the injector's reply topic. Latencies are measured by the injector based on the reply message time. I.e., we compute the \emph{end-to-end} latency since we send the message to Kafka, until the moment we consume from Kafka the aggregation response. These latencies are corrected to take into account the coordination omission problem~\cite{CoordOmi}.
We use two Kafka topics -- one to publish events with 10 partitions; and another to consume responses with 1 partition. Since we only use one Kafka node, replication is set to 1. 
All experiment runs are of 35 minutes, where the first 5 minutes are for warmup, and ignored for latency purposes. We use a fraud dataset from one of our clients to simulate real-world dictionary cardinality for aggregation states.

\subsection{Comparing Flink with Railgun}
\label{subsection:flinkVsRailgun}
\begin{figure}[h]
  \includegraphics[width=0.45\textwidth]{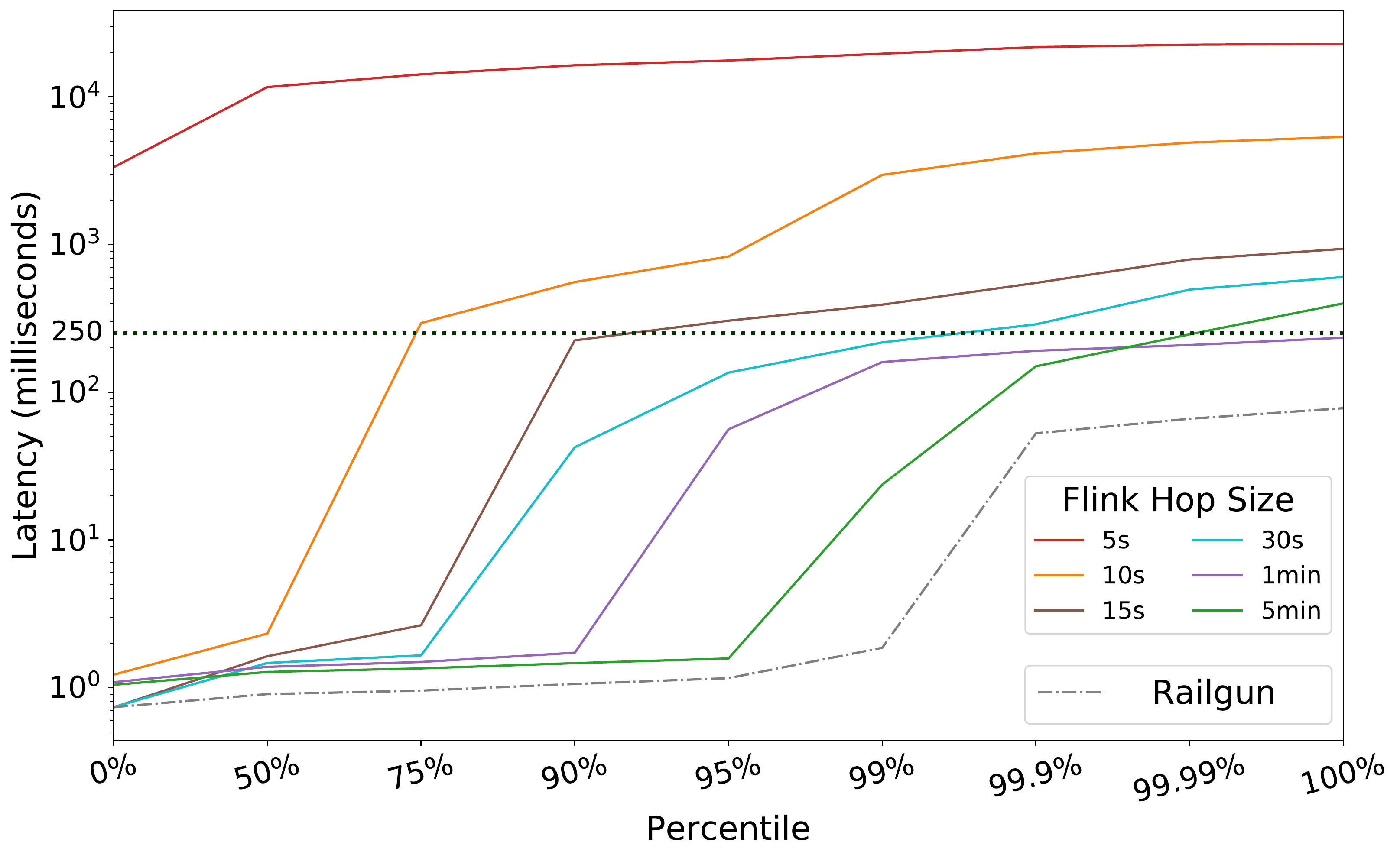}
  \caption{Latency of Flink's hopping windows vs. Railgun's real sliding windows at a fixed throughput of 500 ev/s.}
  \label{fig:flink_vs_railgun}
\end{figure}

Here, we show Flink's architectural limitations stemming from using hopping windows, as described in Section~\ref{subsec:stateofart}, and how they compare with Railgun's real-time sliding window. %
For that, we use Flink to compute the \verb|sum(amount)| per card, over a 60-min hopping window. Then, we vary the hop size from 5 minutes to 1 second to show how Flink latencies behave. To optimize Flink for latency rather than throughput, we set Flink's Kafka Client to use a batch timeout of 0.
The results are shown in Figure~\ref{fig:flink_vs_railgun}, where we include Railgun's latency for the same query using its real sliding window.

\subsubsection{Discussion}
Figure~\ref{fig:flink_vs_railgun} shows how Flink latencies are affected when we increase the hop's granularity. Clearly, in this setup, with hops of 10s or less, Flink is unable to keep with a 500 ev/s throughput.
%
%
Yet, in most of our setups, we are required to score events in less than 250ms in the 99.9\% percentile (cf. \textbf{L} requirement of Section~\ref{sec:realtimewindows}). For those, we need hops of at least 1 minute, which would severely compromise accuracy, and violate rules for our clients (cf. \textbf{A} requirement of Section~\ref{sec:realtimewindows}).
%
Railgun achieves all LAD requirements with lower latencies than Flink, on all percentiles.

\subsection{Scaling Railgun}
\label{subsec:scalingRailgun}
\begin{figure}
    \centering
    \subfloat[Vary Window Size]{{\includegraphics[width=0.45\textwidth]{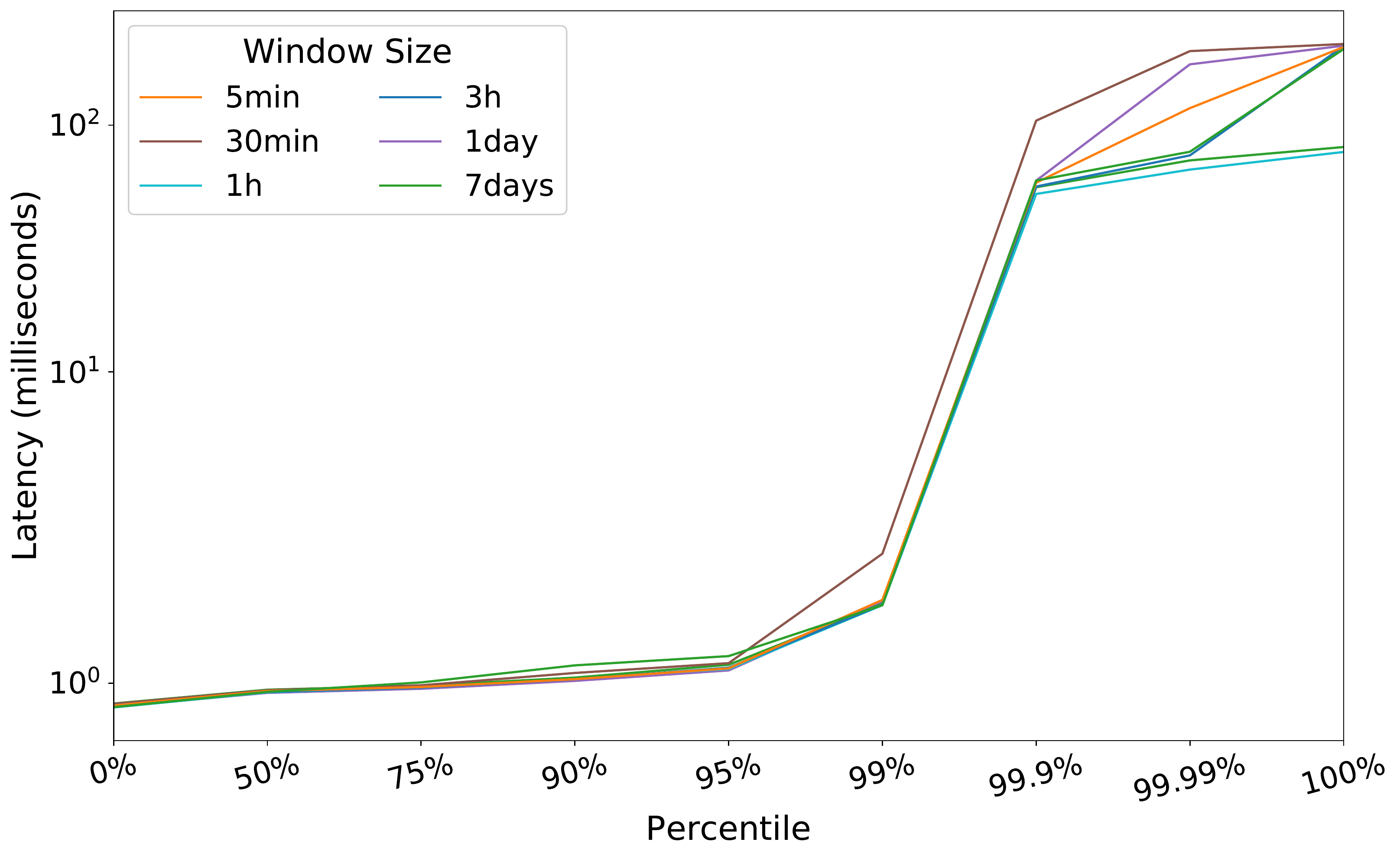} }}%
    \qquad
    \subfloat[Vary Number of Iterators]{{\includegraphics[width=0.45\textwidth]{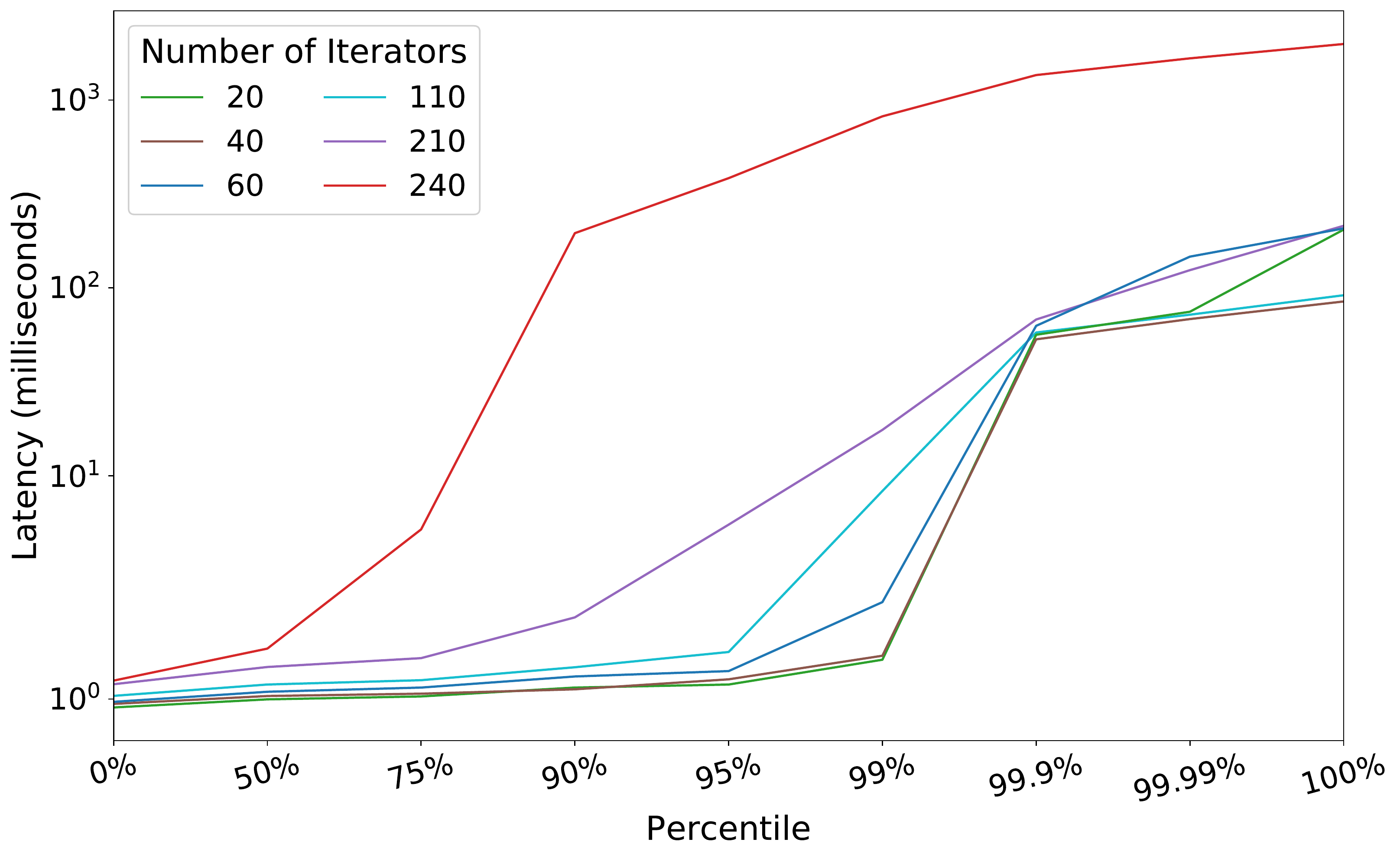} }}%
    \caption{Scaling Railgun}%
    \label{fig:scaling-railgun}%
\end{figure}

To demonstrate how Railgun scales within a single machine, we designed two different experiments. In the first experiment, we compute the same metric as in Section~\ref{subsection:flinkVsRailgun}, but vary the window size from 5 minutes to 7 days. 
In the second experiment, we compute three different metrics: \texttt{sum, average} and \texttt{count}, over the \texttt{amount} field grouped by \texttt{card}. Then, we vary the number of windows on which we compute these three metrics, to enforce a different number of reservoir iterators. Recall from Section~\ref{subsec:reservoir} that the number of iterators depends on the number of windows and how they are aligned. 
When two windows are aligned, either at the beginning or the end, they share the same iterator. Thus, to vary from 20 to 240 iterators, we vary from 10 to 120 misaligned windows. 
The results from both experiments can be seen in Figure~\ref{fig:scaling-railgun}.

\subsubsection{Discussion}
On the first experiment, we clearly show that the window size is irrelevant to Railgun's latency performance. This is expected since for any window we have two iterators, independently of the window size. 
Additional benchmarks have shown us that variations in higher percentiles (>99.99\%), are due to Kafka communication, rather than Railgun (or Flink in Figure~\ref{fig:flink_vs_railgun}). Hence, in some runs we have ~150ms in 99.99\% percentile, and in others ~75ms. 

For the second experiment, we show that as long as the iterators can retrieve the next chunk from cache, the impact on latencies is almost irrelevant. Each iterator requires a chunk in-memory, and, in this experiment, we used 220 chunk elements in Railgun's cache. This means that for most experiments, whenever the iterator requests the next chunk, it is already pre-loaded in memory. Hence, we only start to see some latency degradation when we have almost the same elements in cache as the number of iterators, i.e., when we increase the probability of a cache-miss.
On the run where we have 240 iterators, we also start to see GC problems due to memory pressure, which then leads to higher latencies. This is as expected since the actual heap usage is very close to the maximum JVM heap. 

\section{Conclusions}
In this paper we propose Railgun, a novel distributed streaming engine that supports real-time sliding windows, while providing crucial non-functional requirements: high throughput, low latency, horizontal scalability and fault tolerance. We believe that Railgun is the first distributed streaming engine able to deliver accurate metrics on strict mission-critical scenarios, such as fraud-detection. 

One of Railgun's enablers is the event reservoir. Since accurate metrics require considering all events, the reservoir efficiently persists them to disk, while fetching chunks of events ahead of time as they are needed by windows. This allows Railgun to support time-windows spanning years with the same memory usage as windows of seconds. 
To reduce storage costs, the reservoir uses cheap local HDDs or network-attached disks, and exploits the events' immutability to aggressively compress and serialize them.

Railgun is still under development, and some work lies ahead to validate some components of our design. Namely, we need to certify that: 1) recovery of a task processor does not severely impact the overall system latency; 2) our chosen data format for the reservoir efficiently supports online schema changes and metrics backfill, i.e., the ability to add a new metric and fill it from old event data.
Although still a prototype, our experiments provide sound promises for Railgun. Particularly, we show that Railgun has lower latencies per event than Flink, even when Flink is configured for a low metric accuracy (5-min hop size). Finally, we show that our performance is unaffected by the window size, and that Railgun scales reasonably well with the number of windows and metrics, as long we are able to prevent I/O calls on the critical path of event processing, by pro-actively loading reservoir chunks into memory, ahead of time.

\bibliographystyle{ACM-Reference-Format}
\bibliography{refs}

\end{document}